\documentclass[11pt,twoside]{article}
\usepackage{ntsha_win}
\usepackage[OT1,T2A]{fontenc}
\usepackage[cp1251]{inputenc}
\usepackage[english,ukrainian]{babel}
\renewenvironment{abstract}{\begin{quotation}
\scriptsize
}{\end{quotation}}
\addto\captionsukrainian{%

}

\chardef\atcode=\catcode`\@\catcode`\@=11
\expandafter\let\expandafter\protect\csname protect\endcsname
\def\allowhyphens{\penalty\@M \hskip\z@skip}
\def\set@low@box#1{\setbox\tw@\hbox{,}\setbox\z@\hbox{#1}%
  \setbox\z@\hbox{\dimen@\ht\z@ \advance\dimen@ -\ht\tw@
      \lower\dimen@\box\z@}%
  \ht\z@\ht\tw@ \dp\z@\dp\tw@}
\def\save@sf@q#1{{\ifhmode \edef\@SF{\spacefactor\the\spacefactor}\else
  \let\@SF\empty \fi \leavevmode #1\@SF}}
\def\@glqq{\save@sf@q{\set@low@box{''\/}\box\z@\kern-.04em\allowhyphens}}
\def\ulq{\protect\@glqq}
\def\@grqq{\save@sf@q{\kern-.07em``\kern.07em}}
\def\urq{\protect\@grqq}
\catcode`\@=\atcode

\newfont{\sssm}{cmss8 scaled \magstephalf}
\newfont{\ssb}{cmssbx10 scaled \magstephalf}
\newcommand{\sst}{\scriptscriptstyle}

\newbox\acccel\setbox\acccel=\hbox{$\hbox{\ssb v}^{\prime\prime}$}
\def\ac{\copy\acccel}

\def\B{{\mbox{\ssb B}}}
\def\c{{\mbox{\ssb c}}}
\def\bs#1{{\mbox{\ssb #1}}}
\def\ms#1{{\sf #1}}

\def\bcdot{\mbox{\boldmath$\cdot$}}
\def\bpartial{\mbox{\boldmath$\partial$}}
\def\bm#1{{\mathbf#1}}
\def\ubm#1{\mbox{\unboldmath$#1$}}

\newbox\abox\setbox\abox=\hbox{\ssb a}
\newbox\ssbox\setbox\ssbox=\hbox{\ssb s}
\newbox\Elocity\setbox\Elocity=\hbox{\ssb v}
\newbox\PrimE\setbox\PrimE=\hbox{{\ssb v}$^{\prime}$}
\newbox\PPrimE\setbox\PPrimE=\hbox{{\ssb p}$^{\prime}$}
\newbox\elocity\setbox\elocity=\hbox{\sssm v}
\newbox\Prime\setbox\Prime=\hbox{{\sssm v}$^{\scriptscriptstyle\prime}$}
\catcode`A=13\def A{\copy\abox}\catcode65=11
\catcode`S=13\def S{\copy\ssbox}\catcode83=11
\catcode`V=13\def V{\copy\Elocity}\catcode86=11
\catcode`W=13\def W{\copy\PrimE}\catcode87=11
\catcode`v=13\def v{{\copy\elocity}}\catcode118=11
\catcode`w=13\def w{{\copy\Prime}}\catcode119=11
\catcode`"=13\let "=\wedge\catcode34=12
 \def\W{{\copy\PrimE}}
 \def\PP{{\copy\PPrimE}}
\def\b{\begingroup\catcode`A=13 \catcode`S=13 \catcode`V=13 \catcode`W=13
\catcode`v=13 \catcode`w=13\catcode`"=13}
\def\e{\endgroup}

\title[Буцім-гамільтонівський опис клясичного спіну]{БУЦІМ-ГАМІЛЬТОНІВСЬКИЙ ОПИС КЛЯСИЧНОГО СПІНУ}
\author[Р.~МАЦЮК]{Роман МАЦЮК}
\date{Інститут прикладних проблем механіки і математики\\
ім.~Я.С.~Підстригача НАН~України\\
вул. Наукова 3-Б, Львів 79601
\footnotetext{PACS numbers 03.30.+p, 45.20.Jj, 45.50.Dd}}
\begin{document}
\setcounter{page}{226}

\maketitle
\hfill \small Редакція отримала статтю 7\textsuperscript{го}~вересня 2000\textsuperscript{го}~року.
\bigskip
\thispagestyle{myheadings}
\markboth{~\hrulefill~Фізичний збірник НТШ~т.4 2001~p.}
     {Фізичний збірник НТШ~т.4 2001~p.~\hrulefill~}
\begin{abstract}
Пропонуємо до уваги читача деяке сімейство функцій Ляґранжа, кожна з яких у
стані породити релятивіське рівняння руху третього ряду за похідними, що
описує динаміку вільної клясичної частки, наділеної моментом обертання. На
цій основі запроваджено узагальнено-гамільтонівський опис релятивіської дзиґи,
узгоджений з додатковою умовою Pirani.
\end{abstract}

\section*{ВСТУП}
Успішна ґеометризація загальної теорії ґравітації переконливо свідчить про
те, що ріманівська ґеометрія є для фізики світу чимось більшим, аніж просто
вигідним інструментом для застосування методів ґлобального нутрішнього
аналізу до динаміки фундаментальних фізичних систем. Останньо визнаними
зістали теж і друг\'і ґеометрії, які служать для опису слабих,
електромаґнетних та сильних взаємодій. Але для цих наступних ґеометрій, що є
основою для теорій калібрувальних (відмітних) полів, важливим пунктом є
апріорне феноменологічне встановлення ґрупи внутрішньої симетрії, що лягає в
основу апарату коваріянтного упохіднення, паралельного перенесення, і,
взагалі, усього апарату, який обслуговує ґеометрію відповідних волокнистих
в'язок. Тільки ріманівська ґеометрія не вимагає впровадження додаткових ґруп
симетрій в основу теорії, бо вона пов'язана самодостатнім чином лише з
топологією конфіґураційного простору динамічної системи, і ні з чим більше.
Але сама чиста псевдоріманівська ґеометрія простору-часу не може описати
інших взаємодій, як тільки ґравітаційні. Ця ґеометрія природнім чином
призначена для встановлення правила ґеодезійного руху безструктурних пробних
часток, які не відчувають інших полів. Цей ґеодезійний рух описується
диференційним рівнянням другого ряду (за похідними). Коли присутні інші поля,
ми звикли вдаватись до ґеометрій з додатковими локальними ґрупами (волокнисті
в'язки) і описувати рух пробних часток ґеодезійними лініями в просторах з
більшою кількістю змінних. Рівняння відповідних ґеодезійних ліній теж є не
вище другого ряду за похідними, як, скажімо, в ґеометріях на кшталт теорії
Калюци-Кляйна. Ці ґеодезійні лінії розширених ґеометрій проєктуються в
світові лінії частки у чотиривимірному просторі-часі, які вже відхиляються
від звичайних ґеодезійних. Чи пам'ятає ґеометрія нашого світу про причину
відхилення? Можна спробувати пояснити таке відхилення від ґеодезійности
іншими причинами, обійшовшись без упровадження додаткових вимірів до
фізичного світу. Наприклад, використовуючи апарат такої
чотиривимірної ґеометрії, де ґеодезійні лінії визначаються диференційним рівнянням
вищого ряду за похідними.

Ми задалися питанням: а які взагалі відхилення від ґеодезійного руху допускає
чотиривимірна динаміка в релятивіському просторі-часі. Відомо, що врахування
ефектів ріжної природи, як-от: випромінення, нелокальність взаємодії,
внутрішня будова пробної частки або наділення її додатковими
характеристиками на кшталт спіну,-- провадить до підвищення порядку похідних
у диференційному рівнянні, яке описує розвиток відповідної моделі. Ми
обмежимо своє завдання розглядом тільки рівнянь третього ряду за похідними,
які б мали ту непересічну властивість, що походили б від якогось варіяційного
завдання (тобто йдеться тільки про варіяційне диференційне
рівняння, яке в теорії варіяційного числення для звичайного інтеґрала дії
називають рівнянням Ойлера-Пуасона).

Історія використання ляґранжіянів з вищими похідними для опису клясичної (не
квантової) динаміки фізичних часток сягає 40\textsuperscript{х}~років і пов'язана з
такими іменами: F.~Bopp (1946), H.~H\"onl (1948), З.~Храпливий,
J.~Weys\-sen\-hoff та A.~Raabe. Останні два починали свою працю над вказаною
тематикою ще у Львові в часі першої совєтської окупації. Сьогодні ляґранжіяни з вищими
похідними можна зустріти в теоріях бозонно-ферміонних перемін під зовнішним
полем (Поляков, Плющай, Нестеренко, Iso et al.), при описі клясичної
крутької частки (Riewe), при узагальненні виразу для просторово-часового
інтервалу в релятивізмі (Caianiello), і при спробах будувати фун\-кцію дії з
ди\-фе\-рен\-цій\-них інваріянтів лінії (Якупов, Лейко, Плющай, Нестеренко,
Scarpetta, Arod\'z).
Аналітичні основи варіяційного числення, разом з упровадженням імпульсів
вищого ряду, закладені Михайлом Остро\-гра\-д\-сь\-ким. Не\-спо\-ді\-ва\-ний ріст за\-ці\-кав\-лен\-ня
серед математичного середовища фор\-маль\-но-ґе\-о\-ме\-т\-рич\-ни\-ми аспектами варіяційного
числення вищого ряду за похідними пов'язаний з виникненням і, почавши від
60\textsuperscript{х}~років, швидким розвитком нових підходів до формального
диференційного числення, поширенням методів зовнішнього диференційного
числення в просторах картанівських диференційних форм, а також із
узагальненням теорії диференційно-ґеометричних лучностей на простори з вищими
похідними (т.~зв. простори струменів, або джетів). Ці останні, додатково
наділені функціоналом дії, отримали назву просторів A.~Kawaguchi, який почав
вивчати їх ґеометричні властивості ще в 30\textsuperscript{х}~роках.
Теорія лучностей вищого ряду розвивалася  Ehresman\textsuperscript{ом}, теорія
об'єктів вищого ряду --- Лаптєвим. На сьогодні відповідна література
розрослася. Тільки для прикладу вкажемо на два джерела \cite{Kawaguchi,Krupkova}.

Так само, як плаский простір із пр\'остими лініями в ролі ґеодезійних є
взірцем для простору Рімана, де ґеодезійні лінії описуються складнішим
рівнянням (але все ще другого ряду за похідними),-- так само простір
Мінковського спеціяльної теорії відносности, але вже з лініями, що є
розв'язками деякого рівняння з {\em вищими} похідними,-- може служити локальним
взірцем для значно складнішої ґеометрії деякого викривленого многовиду,
наділеного структурою таких шляхів, що є екстремалями певного функціоналу з вищими
похідними. Такий простір (простір Kawaguchi) відображатиме поведінку частки,
яка зазна\'є додаткових впливів --- байдуже, чи з причини складнішої
внутрішньої будови, під дією зовнішних полів, чи завдяки приписуванню їй
додаткових внутрішніх ступенів вільности спінорного типу.

Обмеживши себе вимогою Пуанкаре-інваріянтности та варіяційности пошукуваних
рівнянь руху не вище третього ряду за похідними, ми одержимо конкретний
вираз, що задасть деяку сім'ю ляґранжіянів. На цій основі збудуємо
узагальнене перетворення Лєжандра і, відповідно до цього, уза\-га\-ль\-не\-но-га\-міль\-то\-нів опис
вільної частки, додатково наділеної вектором клясичного (не квантового) спіну
(т.~зв. {\it крутької\/} частки, {\it крутька,} або ж {\it дзиґи}).

Наш опис узгоджується з т.~зв. додатковою умовою Pirani і може вважатися за
альтернативний до опису \cite{Regge}, котрий, на відміну від нашого, узгоджується з додатковою умовою
Tulczyjew\textsuperscript{а}-Dixon\textsuperscript{а}.

\section{ЛЯҐРАНЖІВ ОПИС}
Відомо\cite{DAN}, що у чотиривимірному просторі марно дошукуватися
варіяційного рівняння третього ряду за похідними, яке мало б відповідну
(псевдо)евклідову симетрію. Зауважимо, що симетрію диференційного рівняння ми
розуміємо в найширшому, але все ще змістовному аспекті: вимагаємо, щоб під
дією відповідних перетворень конфіґураційного простору системи розв'язки
рівняння не сміли переходити у щось иньше, як тільки знову ж у розв'язки
(можливо, другі) того ж рівняння. Відомо також\cite{DAN}, що можна все-ж
віднайти деякі варіяційні рівняння третього ряду з вказаною симетрією,
якщо дозволити додаткову залежність змінних від деякого постійного
чотири-векторного параметра

\[\bm s=(s^{\sst0},\b S\e),\]
який перетворюється за виказом дії
(псевдо)евклідівської групи, але не підлягає варіяції в процедурі виведення
рівняння Ойлера-Пуасона. Загальний вигляд варіяційного рівняння третього
ряду, яке б описувало (непараметризовану) світову лінію пробної частки, є
таким:
\begin{equation}\label{geneq}
\b
\ac\times A+(W\bm.\,\bpartial_v)\;W\times A+\BW+\c=\bs0,
\e
\end{equation}
де векторні функції \b A\e, {\c} та матрична функція {\B} залежать від часу
$t$, просторової координати $\bs x=(\ms x^a)$, швидкості $\bs v=\bs
x^{\prime}=(\ms v^{a})$, компонент чотири-векторного параметра $\bm
s=(s^{\mu})=(s^{\sst0},\ms s^{a})$, і підлягають додатковій системі
диференційних рівняь в часткових похідних, яка забезпечує варіяційну природу
рівняння (\ref{geneq}) \cite{DAN}. Умови Пуанкаре-інваріянтности рівняння
(\ref{geneq}) теж виражаються певною системою рівнянь з частковими похідними, яким
підлягають функції \b A\e, {\c} та {\B} \cite{DAN,thesis}. В роботі \cite{thesis}
було показано, що
сукупність перелічених умов вимагає колінеарности вектора {\b A\e}
і вектора зміщеної швидкости $\bs z=(\bs s-s^{\sst0}\bs v):$
\[\bs a=f(\bs v)\;\bs z.\]
Там же запропоновано наступні вирази для функцій
\b A\e, {\c} та {\B}, які задовольняють усі необхідні вимоги варіяційності та
Пуанкаре-не\-з\-мін\-но\-с\-ти рівняння (\ref{geneq}) (метрику у просторі Мінковського
умовляємось вибрати так, щоб просторова частина діяґонального тензора
$\eta_{\mu\nu}$ дорівнювала одиничній матриці з від'ємним знаком,
$\eta_{\mu\nu}={\rm diag}(1,-1,-1,-1)$):
\begin{eqnarray}
f&=&\b\big[(1+V^{\bm 2})(s_{\sst0}^{2}+S^{\bm 2})-(s_{\sst0}
+S\bcdot V)^{2}\big]^{-3/2}\e\label{f}\\
B_{ab}&=&M_{0}\frac{
(1+\bs v^{\bm 2})\eta_{ab}
-\ms v_{a}\ms v_{b}}{
(1+\bs v^{\bm 2})^{3/2}\;(s_{\sst0}^{2}
+\bs s^{\bm 2})^{3/2}}\label{B}\\
\c&=&\bs 0\label{c}
\end{eqnarray}

Задля того, щоб записати вирази для сім'ї ляґранжіянів, впровадимо такі
позначення. Орти бази чотиривимірного простору-часу позначимо $\bm e_{(\mu)}$,
так що $\bm e_{(\sst0)}\bcdot\,\bm e_{(\sst0)}=\eta_{\sst00}=1$, $\bs e_{(a)}\bcdot\,\bs e_{(a)}=\eta_{aa}=-1$.
Запровадимо допоміжні вектори:
\begin{eqnarray*}
\bs s_{(a)}&=&\bs s-\ms s_{a}\bs e_{(a)}\\
\bs z_{(a)}&=&\bs z-\ms z_{a}\bs e_{(a)},
\end{eqnarray*}
В цих позначеннях шукана сім'я ляґранжіянів запишеться ось як:
\begin{eqnarray}
\lefteqn{L_{a}=\frac{M_{0}}{(s_{\sst0}^{2}+\bs s^{\bm 2})^{3/2}}\sqrt{1+\bs v^{\bm 2}}}\nonumber\\
&&-\frac{s_{\sst0}}{s_{\sst0}^{2}+\bs s^{\bm 2}}\;\frac{(s_{\sst0}^{2}+\bs s_{(a)}^{\bm 2})\,\ms z_{a}
-(\bs s_{(a)}\bcdot\,\bs z_{(a)})\,s_{a}
}{
(s_{\sst0}^{2}+\bs s_{(a)}^{\bm 2})\,\bs z_{(a)}^{\bm 2}
-(\bs s_{(a)}\bcdot\,\bs z_{(a)})^{2}}\;
\frac{\big[\W,\bs z,\bs e_{(a)}\big]
}{
\bs z^{\bm 2}+(\bs s\times\bs v)^{\bm 2}}\label{La}
\end{eqnarray}
Кожна з функцій Ляґранжа $L_{a}$, взята окремо, продукуватиме рівняння (\ref{geneq})
з коефіцієнтами (\ref{f},\ref{B},\ref{c}).

Послужившись простим рецептом \cite{DAN} переходу до одно\-рід\-но-чо\-ти\-ри\-ви\-мір\-ного запису
рівняння (\ref{geneq}), добудемо наступну еквівалентну форму ва\-рія\-цій\-но\-го рівняння третього
ряду для опису
непараметризованих світових ліній крутьких часток:
\begin{eqnarray}
\lefteqn{\frac{M_{0}}{\|{\bm s}\|^3}\left[
\frac{{\bf\dot{\bm u}}{{\bcdot}}\bm u}{\|\bm u\|^3}\,\bm u
-\frac{{\bf\dot{\bm u}}}{\|\bm u\|}
\right]}\nonumber
\\[2\jot]
&&\;-\;\frac{\ast\,{\bf\ddot{\bm u}}\wedge\bm u\wedge\bm s}
	{\|\bm s\wedge\bm u\|^3}
\;+\;3\,\frac{\ast\,{\bf\dot{\bm u}}\wedge\bm u\wedge\bm s}
	{\|\bm s\wedge\bm u\|^5}\,({\bf\dot{\bm u}}\wedge\bm s){{\bcdot}}(\bm u\wedge\bm s)
\,=\,0\,.\label{myeq}
\end{eqnarray}
Просторова частина чотири-вектора (неунітарної, взагалі кажучи) швидкості $\bm u$
збігається з тривимірним вектором $\bs v$ при спеціяльному виборі параметризації
світової лінії, коли $u^{\sst0}=1$.

\section{ГАМІЛЬТОНІВ ОПИС}
Рецепт узагальненого методу Гамільтона-Остроградського запропонований у \cite{Krupkova}.
Поруч із ляґранжевою системою (\ref{geneq}) пропонується поставити під розгляд
узагальнену гамільтонову систему, яка описується ядром замкнутої зовнішньої
диференційної форми
\begin{equation}\label{Pfaff}
-d(\ms p_{a}\ms v^{a}-L_{0})\wedge dt+d\ms p_{a}\wedge d\ms x^{a}+d\ms p'_{a}\wedge d\ms v^{a},
\end{equation}
де $L_{0}$ є неоднорідною решткою від повної функції Ляґранжа у випадку, коли рівняння
Ойлера-Пуасона для повної функції Ляґранжа $L$ не містить похідних четвертого порядку:

\[
L_{0}(t,\bs x,\bs v)\stackrel{\rm def}{=}L(t,\bs x,\bs v,\W)-\W\bm.\,\frac{\partial L}{\bpartial \W}\,.
\]
Узагальнене перетворення Лєжандра задається наступним правилом:

\[
\PP=\frac{\partial L}{\bpartial\W},\qquad \bs p=\frac{\partial L}{\bpartial\bs v}
-\frac{d\,\PP}{dt}.
\]

Щоб записати вираз для узагальненого перетворення Лєжандра,
запровадимо стовпець $\bm\zeta=(\zeta_{a})$:

\[
\zeta_{a}\;=\;\frac1{s_{\sst0}}\,\frac{
(s_{\sst0}+\bs s\bcdot\bs v)\,\ms s_{a}-(\ms s_{\sst0}^{2}+\bs s^{\bm 2})\,\ms v_{a}
}{
\bs z^{\bm 2}-\ms z_{a}^{2}+(\bs s\times\bs v)^{\bm 2}}\,.
\]

Кожен з ляґранжіянів (\ref{La}) продукує одне і те саме перетворення Лєжандра
і, відповідно до цього, однакову функцію Гамільтона $H=\bs p\bm.\bs v+\PP\bm.\W-L$:
\begin{eqnarray}
\bs p&=&\frac{M_{0}}{(s_{\sst0}^{2}+\bs s^{\bm 2})^{3/2}}\;
      \frac{\bs v}{\sqrt{1+\bs v^{\bm 2}}}
     +\frac{\W\times\bs z}{\big[\bs z^{\bm 2}+(\bs s\times\bs v)^{\bm 2}\big]^{3/2}}\label{p}\\
  \PP&=&\frac{\zeta\times\bs z}{3(s_{\sst0}^{2}+\bs s^{\bm 2})\sqrt{\bs z^{\bm 2}+(\bs s\times\bs v)^{\bm 2}}}\label{PP}\\
    H&=&-\frac{M_{0}}{(s_{\sst0}^{2}+\bs s^{\bm 2})^{3/2}\sqrt{1+\bs v^{\bm 2}}}
        -\frac{[\W,\bs v,\bs s]}{\big[\bs z^{\bm 2}+(\bs s\times\bs v)^{\bm 2}\big]^{3/2}}.\label{H}
\end{eqnarray}

\section{УЗАГАЛЬНЕННЯ УМОВИ PIRANI ТА СПІНОРНА ЗАЛЕЖНІСТЬ ВЛАСНОЇ МАСИ}
Узагальнена система Гамільтона-Остроградського (\ref{Pfaff}), яка відповідає
узагальненому перетворенню Лєжандра (\ref{p},\ref{PP}) з функцією Гамільтона
(\ref{H}), має в собі (за термінологією \cite{Krupkova}) первинну бросткову в'язь
(primary semispray constraint):
\begin{equation}\label{Constr}
\frac{M_{0}}{s_{\sst0}^{2}+\bs s^{\bm 2}}\;\left[\frac{\bs s\bcdot\bs\W}{\sqrt{1+\bs v^{\bm 2}}}
-\frac{(\bs v\bcdot\W)(s_{\sst0}+\bs s\bcdot\bs v)}{(1+\bs v^{\bm 2})^{3/2}}\right]
=0.
\end{equation}
Можна рівнозначно висловитись, що рівняння (\ref{geneq}), або, еквівалентно,
рівняння (\ref{myeq}) має першого інтеґрала:
\begin{equation}\label{int}
\frac{s_{\sst0}+\bs s\bcdot\bs v}{\sqrt{1+\bs v^{\bm 2}}}=K.
\end{equation}
При значенні $K=0$ рівність (\ref{int}) є нічим иньшим, як в'яззю \cite{DAN}
\boldmath
\begin{equation}\label{pirani}
s\cdot u=\ubm0\,,
\end{equation}\unboldmath
котра накладається
відомою умовою Pirani

\[
u_{\mu}S^{\mu\nu}=0,\qquad \mu,\nu=0,1,2,3\,,
\]
де тензор $S^{\mu\nu}$ та коваріянтний чотири-вектор $\bm s=(s_{\sst0}, s_{\mu})$
пов'язані співвідношенням

\[
s_{\nu}=\frac1{2\|\bm u\|}\epsilon_{\kappa\mu\rho\nu}u^{\kappa}S^{\mu\rho}\,.
\]
Тому рівність (\ref{Constr}) назвемо узагальненою умовою
Pirani.

В роботах \cite{DAN,thesis} нами показано, що рівняння (\ref{myeq}) робиться еквівалентним
до рівнянь Mathisson\textsuperscript{а}-Papapetrou \cite{Math} з додатковою умовою Pirani
у двох випадках: 1)~або ми обмежуємо вибір інтеґрала руху (\ref{int}) значенням
$K=0$, або 2)~переходимо до нового значення власної маси частки, яке теж є постійним
уздовж траєкторії руху в силу самого рівняння (\ref{myeq}):
\begin{equation}\label{mass}
m_{0}=M_{0}\left[1-\frac{(s_{\sst0}+\bs s\bcdot\bs v)^{2}}{(s_{\sst0}^{2}+\bs s^{\bm 2})
(1+\bs v^{\bm 2})}\right]^{3/2}.
\end{equation}

\section*{ОБГОВОРЕННЯ}
\paragraph*{\bf1.}
Найперше ми були зацікавлені у побудові диференційної форми (\ref{Pfaff}),
і для цього необхідно було здогадатись, як повинні виглядати формули (\ref{p}) і
(\ref{PP}). Функція $L_{0}$ від другої похідної не залежить, тому вона не
вважається центральним місцем наших рахунків, як і вираз (\ref{H}), що його
найцікавішою частиною є другий доданок, тобто згортка $\bs p\bm.\bs v$.
Форму (\ref{Pfaff}) можна виражати в різноманітних координатах. У виразі
(\ref{Pfaff}) навіть не всі диференціяли є незалежними. Ми не ставили
поки що завдання про обернення формул (\ref{p}) і
(\ref{PP}). Відповідно до термінології \cite{Krupkova}, йде мова про т.~зв.
{\it узагальнене} перетворення Лєжандра і про подання форми (\ref{Pfaff}) у
т.~зв. {\it узагальнених лєжандрівських} координатах. Але ми назвали запропонований опис
вертлявої частки {\it буцім-гамільтонівським} з іще поважнішої причини: не
всі інтеґральні многовиди ядра замкненої форми (\ref{Pfaff}) є
автоматично голономними (як це є у клясичному гамільтонівському описі). Тому
{\ulq}узагальнено-гамільтонівська{\urq} динаміка, що описується ядром форми
(\ref{Pfaff}), відрізнятиметься від ляґранжівської там, де не діє первинна
бросткова в'язь (\ref{Constr}). Окрім цього, не обговорювалися ні векторна
природа, ні ґеометрична структура фазового простору, утвореного узагальненими
імпульсами.
\paragraph*{\bf2.}
Для опису руху часток з нутрішнім обертовим моментом
прий\-нято застосовувати диференційні рівняння Papapetrou \cite{wos,pap}, які,
річ ясна, не містять похідних від прискорення. До цих рівнянь часом долучають
додаткову умову Pirani. Нам невідомо, щоб умова Pirani, вкупі з рівняннями
Papapetrou, {\it прямо} випливала з якогось варіяційного принципу --- без
апріорних в'язей, або без додаткових постеріорних обмежень фізичного чи
математичного характеру. З иньшого боку, перше диференційне продовження
\cite{thesis} динамічної системи, утвореної рівняннями Papapetrou вкупі з
умовою Pirani, дозволяє запропонувати таке варіяційне узагальнення, при якому
{\ulq}узагальнена умова Pirani\urq (\ref{Constr}) є {\it наслідком}
варіяційних рівнянь (\ref{myeq}).

Уперше рівняння третього ряду за похідними використовувалось у явному вигляді
для опису крутьких часток у праці Mathisson\textsuperscript{а} \cite{Math}, однак
питання про запис функції Ляґранжа для нього у вказаній праці не
обговорювалось. Принцип варіяційности для рівнянь третього ряду
наштовхує ось на яке вирішення:
\begin{enumerate} \item
Пропонується узагальнити додаткову умову Pirani таким чином, щоб вона
випливала з варіяційного завдання для функції дії з таким ляґранжіяном, який
містить похідні другого порядку від координат частки.
\item Поруч із
частками, для яких виконується {\ulq}умова ор\-то\-го\-на\-ль\-но\-с\-ти{\urq}
(\ref{pirani}), розглядаються також і частки, світові лінії яких
нахилені під певним (постійним) кутом до напряму чотири-вектора
спіну,

\boldmath\[
\frac{s\cdot u}{\|u\|}=\mbox{const}\,.
\]\unboldmath
\end{enumerate}

\bigskip
\begin{center}
\centerline{\textbf{QUASI-HAMILTONIAN DESCRIPTION OF CLASSICAL SPIN}}
\medskip
\emph{Roman MATSYUK}
\\
\medskip
Pidstryhach Institute for Applied Problems\\
in Mechanics and Mathematics\\
of the National Academy of Sciences of Ukraine\\
3\textsuperscript{b}~Naukova~St., Lviv, 79601, Ukraine
\end{center}
\bigskip
A family of Lagrange functions is considered, each producing the classical
relativistic free spinning particle equation of motion of the third
order. On this grounds a generalized Hamilton-Ostrohrads'kyj description of
the free relativistic spherical top is proposed, which comply with the
Pirani supplementary conditions.
\end{document}